\documentclass[iop]{emulateapj}
\usepackage{graphicx}
\usepackage{epstopdf}
\usepackage{natbib}
\usepackage{xspace}
\usepackage{color}

\newcommand{\iras}   	{IRAS~16293$-$2422\xspace}
\newcommand{\irasA}   	{I16293A\xspace}
\newcommand{\irasB}   	{I16293B\xspace}

\newcommand{\ie}    	{i.e.,}
\newcommand{\eg}    	{e.g.,}
\newcommand{\ruv}    	{$r_{\rm u,v}$}
\newcommand{\kl}    	{k$\lambda$}
\newcommand{\mJyB}	{mJy~Beam$^{-1}$}

\begin{document}

\title{Detection of a Magnetized Disk around a Very Young Protostar}

\author{Ramprasad Rao\altaffilmark{1}, 
Josep M. Girart\altaffilmark{2}, Shih-Ping Lai\altaffilmark{3,4}, 
Daniel P. Marrone\altaffilmark{5}
}

\affil{$^1$Institute of Astronomy and Astrophysics, Academia Sinica, 645
N. Aohoku Pl., Hilo, HI 96720, USA}
\affil{$^2$ Institut de Ci\`encies de l'Espai, (CSIC-IEEC), Campus UAB,
Facultat de Ci\`encies, C5p 2, 08193 Bellaterra, Catalonia;
girart@ice.cat}
\affil{$^3$Institute of Astronomy and Department of Physics, National
Tsing Hua University, Hsinchu 30013, Taiwan}
\affil{$^4$Academia Sinica Institute of Astronomy and Astrophysics, P.O. Box
23-141, Taipei 10617, Taiwan}
\affil{$^5$Steward Observatory, University of Arizona, 933 North Cherry Avenue, 
Tucson, AZ 85721, USA}

\begin{abstract}
We present subarcsecond resolution polarimetric observations of the 878~$\mu$m 
thermal dust continuum emission obtained with the Submillimeter Array (SMA)  towards 
the IRAS~16293$-$2422 protostellar  binary system. We report the detection of 
linearly polarized dust emission arising from the circumstellar disk 
associated with the \iras\ B protostar. The fractional polarization of  
$\simeq 1.4$\%  is only slightly lower than that expected from  theoretical 
calculations in such disks.  The magnetic field structure on the plane of the sky 
derived from the dust polarization suggests a complex magnetic field geometry 
in the disk, possibly associated with a rotating disk that is wrapping the field lines 
as expected from the simulations. The polarization around \iras\ A at 
sub-arcsecond angular resolution  is only marginally detected. 
\end{abstract}


\section{Introduction}\label{intro}

The protostellar phase of star formation in a low mass star  is thought to have the 
following structure: a deeply embedded protostar which is not observed directly, 
accompanied by an accretion (protoplanetary) disk, from where a powerful bipolar 
outflow is ejected, and further surrounded by a dense and warm envelope \citep{Andre00}.  
The existence of rotationally supported disks in the earliest stages of the protostellar 
phase (Class 0 protostars) is a matter of debate \citep{Jorgensen09, Girart09b, Tobin13, 
Yen13, Murillo13}, since it is believed that magnetic braking can efficiently suppress 
the formation of a disk in the protostellar phase \citep[\eg][]{Li11}.

\iras\ (henceforth I16293) is  located in the $\rho$ Ophiuchus cloud at a distance of 
120~pc \citep[e.g.,][]{Loinard08}. It is one of the most extensively studied low mass 
protostellar objects and  powers a multiple outflow system \citep{Walker88, Mizuno90, 
rao09, Loinard13}.  At scales of a few hundred AU,  the molecular emission shows 
two dense and  warm condensations, sources \iras\ A (henceforth \irasA) and \iras\ B 
(henceforth \irasB), with a very rich chemical  composition \citep{Chandler05, 
Takakuwa07, Zapata13}.   Very Large Array (VLA) observations at centimeter (cm) 
wavelengths show that \irasA has complex structure in the  free-free emission  
\citep{Wootten89,Mundy92, Chandler05}, whereas the dust emission appears to be 
significant even  at 1.3 cm for \irasB \citep{Estalella91, Loinard07}.  VLA observations 
with an angular  resolution of $0\farcs07$ at 7~mm show that the emission in \irasB\ 
arises from the dust and is well resolved, showing a clear disk--like morphology, which 
is possibly face--on \citep{Rodriguez05, Chandler05, Loinard13}. \irasA is composed 
of two compact sources, A1 and A2, separated by $\sim$$0\farcs35$, surrounded 
by extended emission, but that emission arises primarily from free--free radiation 
\citep{Pech10, Loinard13}.

Aperture synthesis observations of the linearly polarized dust continuum
emission at mm and submm wavelengths have been successful in revealing the magnetic
field properties in dense cores at scales from few hundred to few thousand AU for
both low and high mass star forming regions \citep{Rao98, Girart99, Girart06,
Girart09, Lai03, Tang09,Hull13}. However, it has been more challenging to look
for  magnetic field signatures at circumstellar scales. Indeed, there have
been a few attempts to detect the dust polarization from circumstellar disks with
mm and submm arrays (CARMA and SMA). However, these have been unsuccessful 
with no positive detections but with stringent upper limits of $\simeq 0.5$\% 
\citep{Hughes09, Hughes13}.  The authors explain that these non-detections are 
possibly due to inefficient grain alignment.

The earliest polarimetric observations toward I16293 were limited by low sensitivity  
\citep{Flett91, Tamura93, Akeson97}.  Previous SMA observations at an angular 
resolution of $\sim$2$\arcsec$ show that the  large scale global direction of the field  
appears to lie along the dust ridge joining the two emission peaks. Yet, while \irasA\ 
is threaded by an ``hourglass''--like magnetic field structure, \irasB, shows a relatively
ordered magnetic field  with no evidence of  deformation \citep{rao09}.  
Circular and linear  polarization have been detected in the 22~GHz water maser
line around \irasA, with a derived total magnetic field strength of 110 mG
\citep{Alves12}. The water maser emission is probably arising in the interaction/shocked
zoned of the  outflow with the dense circumstellar gas.

In this letter, we present higher resolution SMA polarimetric observations of the continuum
emission at 345 GHz toward I16293.  The  CO 3--2, SiO 8--7, C$^{34}$S 7--6
molecular  line data of these observations have already been presented in a
different paper  \citep{Girart13}, showing that at scales of few hundreds AU the
outflow activity  is centered in  \irasA, with no clear indication of active
outflow activity  toward \irasB.  Here, we report the first detection of the
linear polarized dust  thermal emission from a circumstellar disk, 
the one associated with \irasB.

\section{Observations}

The SMA  observations were taken on 2010 August 28 in the extended configuration.
The receiver was tuned to cover the 333.5-337.5~GHz and  345.5-349.5~GHz
frequencies in the lower (LSB) and upper side band, (USB) respectively. 
The phase center of the telescope was RA(J2000.0)$=16^{\rm h}23^{\rm m}22\fs90$ 
and DEC(J2000.0)$= -24\degr28\arcmin 35\farcs73$ and all the maps presented 
in this paper are centered at this position.   The gain calibrator was the QSO 
J1733-130. The bandpass and polarization calibrator was 3c454.3. The absolute 
flux scale was determined from observations of Neptune, with a flux uncertainty 
of $\sim20$\%.   The data were reduced using the MIRIAD software package.   
The details  of the polarization techniques and calibrations employed 
at the SMA are discussed in \citet{Marrone08} and \citet{Marrone06}.   The instrumental 
polarization or ``leakages'' were found to be between 1 and 2\% for the USB, while 
the LSB leakages were between 2 and 4\%. These leakages were measured to an accuracy 
of 0.1\%.  Self-calibration was performed independently for the USB and LSB on the
continuum emission from \iras.  This data from the  extended configuration was
combined with the data obtained in the compact  configuration taken in April
2006, taking into account the proper motion of this region  \citep{Loinard07}. 
The results from the previous compact configuration observations  have been 
reported by \citet{rao09}. The combination resulted in maps  with both greater 
sensitivity and greater dynamic range. All the maps were generated using 
natural weighting, which maximizes the sensitivity.

\begin{figure*}[h]
\epsscale{1.0}
\includegraphics[width=6.2cm,angle=-90]{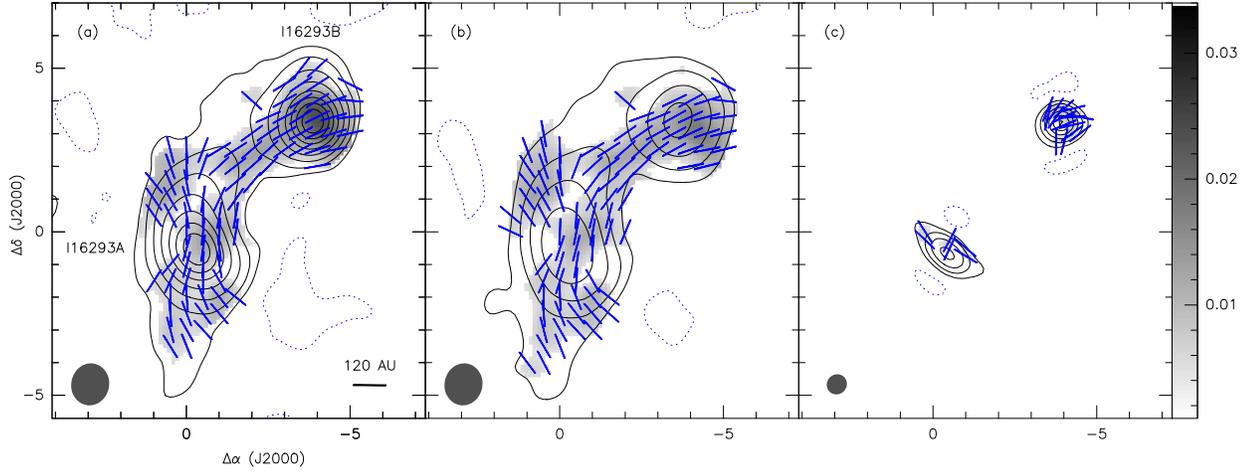}
\caption{
Composite of the contour maps of the 878~$\mu$m dust continuum 
emission, overlaid with the greyscale images of the polarized dust intensity
and the magnetic field directions on the plane of the sky (blue
segments) as derived from the polarized data.
{\it Panel $a$}: Maps obtained using the combined SMA compact and 
extended configurations. 
{\it Panel $b$}: Maps for only the envelope contribution (obtained after 
subtracting the clean components of the map shown in panel $c$, 
see Section~\ref{res} for more  details). 
{\it Panel $c$}: Maps obtained by using only baselines between 80 and 220~\kl. 
For all the panels the contour levels are $-0.08$, 0.08, 0.2, 0.4, 0.7, 1.1,
1.5, 1.9 and 2.3 ~Jy~Beam$^{-1}$ and the grey scale is the same.  The wedge on
the right of the $c$ panel shows the polarized intensity scale in
Jy~Beam$^{-1}$.  The synthesized beams are shown in the bottom-left corner of
each panel: The beam's full width at half maximum (FWHM) and its position angle are
$1\farcs29\times1\farcs16$ and $-10.8\arcdeg$ for the $a$ and $b$ panels,  
and  $0\farcs63\times0\farcs61$ and $-40.0\arcdeg$ for the $c$ panel (these 
resolutions imply a spatial resolution of 146~AU and 75~AU, respectively).
}
\label{Fig1}
\end{figure*}

\begin{figure}[h]
\epsscale{1.0}
\includegraphics[width=7.3cm]{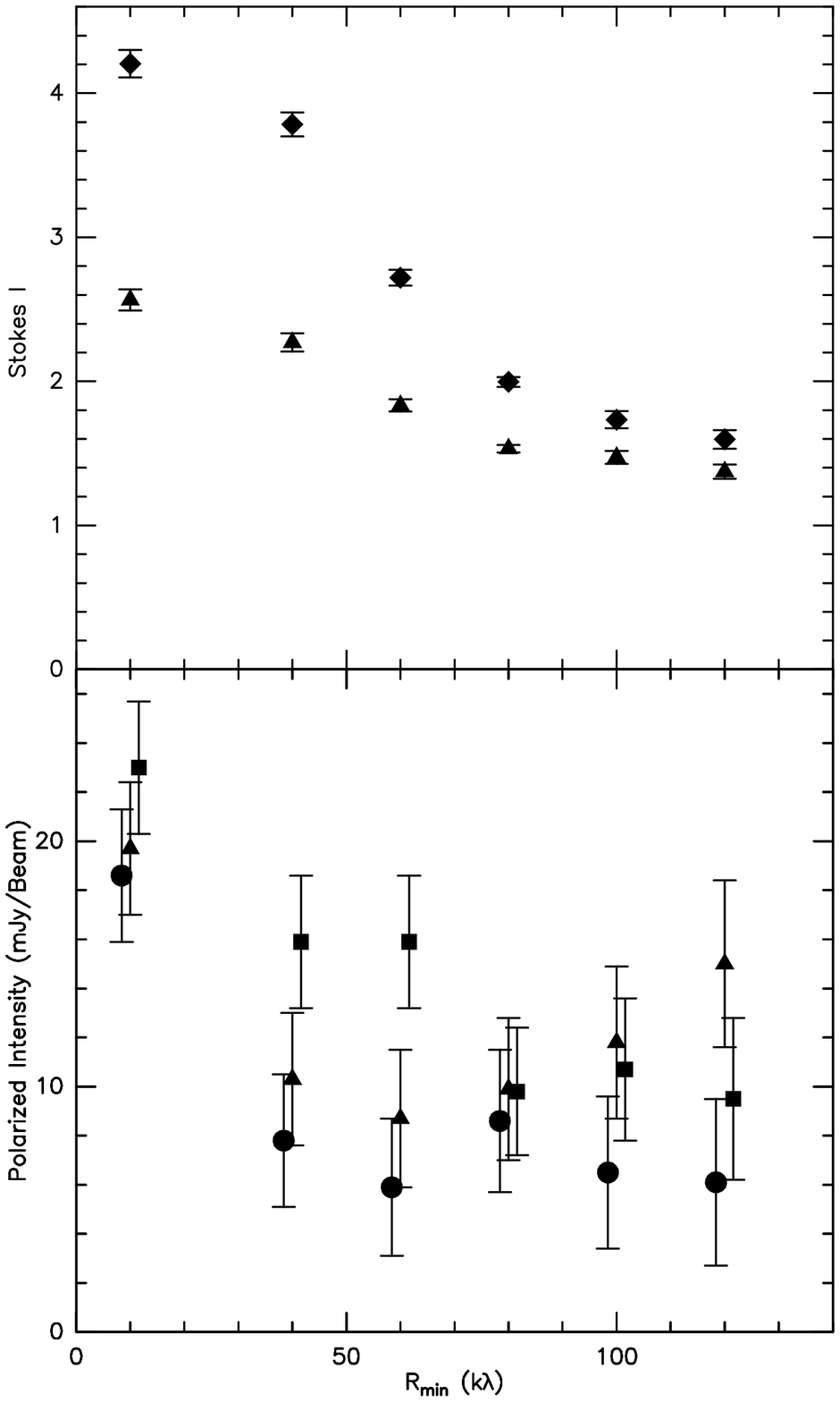}
\caption{
{\it Panel $a$: } Dust continuum 878~$\mu$m emission around \irasB as a 
function of  \ruv\ in units of \kl\ (see Section~\ref{res} for the definition of this
parameter).  Filled rhombus and triangle show the total flux  and peak intensity,
respectively. 
{\it Panel $b$: } Polarization intensity for 3 different positions in \irasB
as a  function of \ruv.  These positions are  marked  in 
Fig.~\ref{Fig3}$a$.
}
\label{Fig2}
\end{figure}

\section{Results and Analysis}\label{res}

The combined 878~$\mu$m dust map and the magnetic field distribution of the  SMA
extended and compact configurations resemble very well the maps reported by 
\citet[][Fig.~\ref{Fig1}$a$ and the bottom panel of their Fig.~2]{rao09},  
in spite of the fact that the combined maps have a smaller beam  ($1\farcs2$) than 
the earlier maps ($2\farcs5$).  The dust continuum emission and the polarization 
detected using the  combined data probably arise both from the envelope
surrounding \irasA\ and \irasB,  as well as from the putative compact, disk-like
structures around these two objects \cite[\eg][]{Rodriguez05, Loinard13}. Henceforth,
we try to approximately isolate the  disk and envelope contributions.  

\begin{figure}[h]
\epsscale{1.0}
\includegraphics[width=9cm]{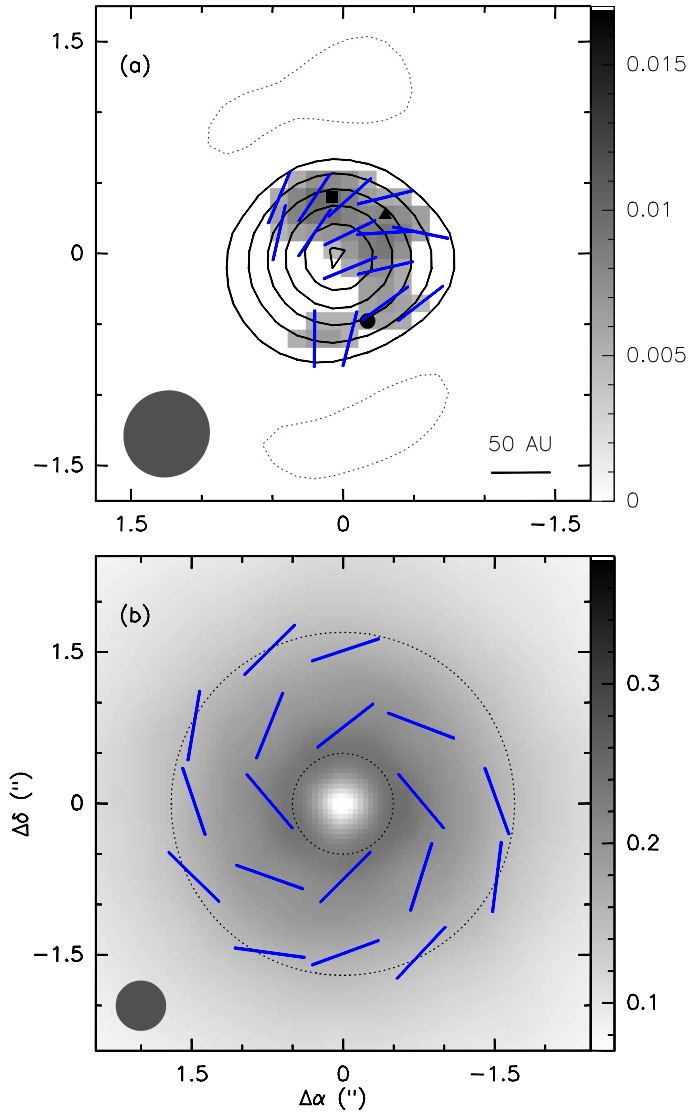}
\caption{
{\it Panel $a$}: A close up of Figure~\ref{Fig1}$c$ toward \irasB.
The filled triangle, square and circle show the selected position shown 
in Figure~\ref{Fig2}$b$.
These positions are separated between $0\farcs40$ and 
$0\farcs76$ (the beam's FWHM is $0\farcs61$).
{\it Panel $b$}: Synthetic map of the linearly polarized dust emission (grey scale),
magnetic field direction (blue segments) and the dust emission (dotted lines). 
This map has been adapted from the Figure~8 of  \citet{Padovani12}. It shows the
magnetic field at circumstellar disk scales derived for a collapsing magnetized dense 
core located at a distance of 140~pc and at time of collapse of $1.94\times10^4$~yr 
\citep{Hennebelle08, Hennebelle09}.
The total and polarized dust emission of the synthetic maps
are in an arbitrary scale  and are shown for qualitative analysis.
}
\label{Fig3}
\end{figure}

In order to trace the putative circumstellar disks and resolve out the contribution 
from the envelope, we obtained maps excluding all the visibilities that have a radius 
lower than \ruv.  We generated maps with different values of \ruv\ (40, 60, 80, 100 
and 120~\kl). The criterion used to  define the radius (\ruv) at which the emission 
arises mostly from circumstellar scales was the minimum value where the flux density 
does not change significantly, i.e. the rate of decrease of flux density with baseline length approaches zero.   
\irasB is an ideal test source since it has an 
accretion disk around the protostar \citep{Rodriguez05, Pineda12, Zapata13}.  
Figure~\ref{Fig2} shows the  peak intensity and  flux density of \irasB for different 
values of \ruv.  The flux and peak intensity decrease with \ruv, but for \ruv$\ga 80$~\kl\ 
the decrease is minimal, indicating that most of the envelope has been resolved out. The 
map obtained with \ruv$=80$~\kl\ is shown in Figure~\ref{Fig1}$c$ (a close-up 
toward \irasB\ is also shown in Fig.~\ref{Fig3}).  To further test that most of the emission 
is associated with the disk, we have measured the source size and its flux.  A Gaussian 
fit to \irasB\ gives a deconvolved size of $0\farcs39 \times 0\farcs28$  (47~AU$\times$34~AU, 
see Table~\ref{tbl}).  These values are consistent with  previous measurements of the 
\irasB\ disk size \citep{Rodriguez05, Zapata13}.  In addition, using the flux derived at 
345~GHz from this map, $1.98\pm0.05$~Jy, and the flux measured from $0\farcs2$ angular 
resolution ALMA observations at 690~GHz \citep[note that at this angular resolution ALMA 
detects only the disk,][]{Loinard13}, $12.5\pm0.5$ Jy, the derived dust continuum spectral 
index is $2.67\pm0.37$.  This  value is in agreement with the previous values measured in 
the centimeter to  submillimeter wavelength range for the disk component \citep{Rodriguez05}.  
Thus, most of the dust emission detected in the  \ruv$=80$~\kl\ map is likely tracing the 
putative circumstellar disk around \irasB.

\begin{deluxetable}{lccc}
\tablecaption{Properties of \irasA and \irasB\label{tbl}}
\tablehead{
\colhead{} & \colhead{Peak intensity} & \colhead{Flux density}
& \colhead{Size \tablenotemark{a}}
\\
\colhead{Component}	& \colhead{(Jy~Beam$^{-1}$)}	& \colhead{(Jy)}
} 
\startdata
All
\irasA\ & $2.27\pm0.04$ & $6.30\pm0.17$ & --- \\
\irasB\ & $2.69\pm0.04$ & $4.47\pm0.24$ & --- \\
\multicolumn{2}{l}{Envelope component}& 
\\
\irasA\ & $1.08\pm0.03$ & $4.80\pm0.18$ & $3\farcs12\times1.56$, 3\arcdeg\\
\irasB\ & $0.84\pm0.03$ & $2.30\pm0.11$ & $1\farcs82\times1.75$, 88\arcdeg \\
\multicolumn{2}{l}{Compact component}
\\
\irasA\ & $0.79\pm0.02$ & $1.57\pm0.07$ & $1\farcs03\times0.10$, 50\arcdeg\\
\irasB\ & $1.56\pm0.02$ & $1.98\pm0.05$ & $0\farcs39\times0.28$, 80\arcdeg \\
\enddata
\tablenotetext{a}{Deconvoled size and position angle}
\end{deluxetable}

The polarized dust emission measured in the \ruv$=80$~\kl\ map, 0.02~Jy, is smaller 
than in the map obtained using all the visibilities, 0.12~Jy. Even though we have 
shown that the total dust emission in the \ruv$=80$~\kl\ map arises from the disk, 
it does not directly imply that the polarized emission arises only from the disk. The 
Stokes $Q$ and $U$ visibility response is quite different from the total emission.  Thus, 
it is possible that the polarized emission may have some contribution from the envelope. 
To qualitatively test the possible envelope's contribution we have generated artificial 
Stokes $Q$ and $U$ maps using a combination of positive and negative Gaussians 
with full width half maximum sizes of $1-4\arcsec$. The criterion used for the artificial 
maps is that the  polarized flux and 
the polarization pattern using all the visibilities is similar to the observed one as is shown 
in Figure~\ref{Fig1}$a$.   
These maps were converted into visibilities using the visibility coverage of our SMA 
observations.  Then, we obtained synthetic SMA maps using all the visibilities, as well as 
using \ruv$=80$ and 120~\kl.   We found that the artificial envelope's polarization 
detected in the \ruv$=80$ and 120~\kl\ maps, 4 and 2~\mJyB\,respectively, is smaller 
than the measured value toward \irasB, 10~\mJyB. Moreover, while the artificial 
envelope's polarized flux clearly decreases between \ruv$=80$ and 120~\kl, the 
observed one remains constant. In  Figure~\ref{Fig2}$b$,
we show, for the different \ruv\ maps, the polarized flux measured at three 
different positions within \irasB. Thus, although we cannot discard definitively the fact
that the 80~\kl\ polarization maps have some residual envelope contribution, it seems 
that most of the polarization detected arises probably from the disk. Interestingly, the 
polarization seems to arise from a remarkable partial ring structure (see 
Figure~\ref{Fig3}). The magnetic field  derived  from the the dust polarization has a 
different morphology  than the one  associated using all the visibilities.

The \ruv$=80$~\kl\ map (Fig.~\ref{Fig1}$c$) shows that the  dust emission 
around \irasA\ is elongated in the NE-SW direction (PA$ \simeq50\arcdeg$) with a 
major axis of $\simeq 120$~AU (see Table~\ref{tbl}).  The direction of the elongation 
is similar to what has been found previously \citep{Loinard13, Girart13}. However, 
the size measured is 
smaller than the rotating molecular structure observed with the C$^{34}$S 7-6 line 
(diameter of 280~AU), which is likely tracing a rotating circumbinary disk around 
sources A1 and A2 \citep{Girart13}.  This suggests that we are resolving partially 
the emission of this structure.  The dust polarization associated with \irasA\ shows only
three weak patches of emission,  the magnetic field direction in two of them is 
along the disk axis and in the other it is perpendicular. However, since the emission is 
patchy and the significance of these detections is more marginal than those toward 
\irasB, we focus the  discussion on \irasB.

To obtain the map with the envelope contribution only, the clean components of the 
\ruv$=80$~\kl\ map (Fig.~\ref{Fig1}$c$) were subtracted from the original 
visibilities for both the compact and extended array data \citep[\eg][]{Frau11}. 
The data  from the two sidebands and from each array configuration were treated 
separately  when doing the subtraction. The map obtained using the resulting visibilities 
is shown in Figure~\ref{Fig1}$b$.  The map showing just the envelope 
contribution  looks similar to the map using all the data, with the difference that 
the envelope emission is considerably smoother (\ie\ not as strongly peaked) in \irasA, 
and significantly more so in  
\irasB.  Table~\ref{tbl} shows the measured flux density, peak intensity and 
size of the envelope contribution. About 76\% and 51\% of the total flux measured 
in  \irasA and \irasB, respectively, arises from the envelope.  The envelope is 
slightly smaller in \irasB\ ($1\farcs8$ or 210~AU) than in \irasA\ ($2\farcs2$ or 
260~AU). The continuum flux of \irasB\  is also smaller, by a factor $\simeq 2$, 
than that of  \irasA. This suggest that if the temperatures in the envelope  are not 
too different, the mass of  the envelope in \irasB\  is a factor 2 smaller than the 
envelope mass in  \irasA\ . Interestingly, the overall pattern of the magnetic  
field of the envelope also 
resembles well the pattern obtained using all the data. In addition, the polarized 
flux from the envelope (0.11~Jy) is similar to that measured using all the data 
(0.12~Jy). This suggests that using all the visibilities from the compact and 
extended configurations, the magnetic field traced is dominated by the contribution 
from the envelope. The contribution to the polarized flux from the disk arises from a much smaller
area when compared to the contribution from the envelope.

\section{Discussion and Conclusions}

In the previous section, we  have shown that the 878~$\mu$m total and polarized dust 
emission map obtained with a \ruv\ of 80~\kl\ (Fig.~\ref{Fig3}$a$) is likely tracing the 
previously reported circumstellar  disk around \irasB.  The magnetic field pattern derived 
from the dust polarization and associated with the disk (Fig.~\ref{Fig1}$c$) appears to 
be distinct  from the field structure as seen in the envelope  (Fig.~\ref{Fig1}$b$). And 
indeed, the magnetic field direction appears to rotate along the ring \ie\ azimuthally. 
Only the vectors in the South-West section of the  ring depart from this general 
trend. This is suggestive of a toroidal magnetic field  that is being wrapped by 
the rotation of the disk
\citep{Hennebelle09, Kataoka12}.  As a qualitative comparison, Figure~\ref{Fig3}$b$ shows the synthetic map of the expected magnetic field pattern
in a rotating circumstellar disk with a  face-on configuration for the case of a
low mass protostar located at  140~pc, and thus is at a distance not too
different from our target, obtained by \citet{Padovani12}.  
Interestingly, the \irasB magnetic field pattern roughly resembles the  
expected configuration for a magnetized rotating, disk.

Theoretical studies of magnetized cores show that it is very difficult to
form a circumstellar disk in the earliest stages of the collapse
\citep[e.g.,][]{Krasnopolsky10, Machida11, Li11}. This is because magnetic braking 
is so effective that only a very small, observationally undetectable, disk can be
formed around the protostar \citep{Dapp12}.  Only at later stages, when the 
surrounding envelope becomes less massive, magnetic braking becomes inefficient, 
allowing a circumstellar disk to grow \citep{Machida11b}. Nevertheless, other 
simulations show that when rotation and magnetic field axis are misaligned, then 
magnetic braking can be inefficient, and for large misalignments a relatively massive 
circumstellar disk is formed in the protostellar phase \citep{Joos12}.  For the case 
of \irasB, the data presented here suggests that this source has a more diffuse 
envelope than \irasA: on one hand the flux density (and therefore likely its mass) of
the \irasB\  envelope is a factor 2 lower than in the \irasA\ envelope; on the other 
hand, 50\% of the flux measured toward \irasB\ using the compact and 
extended configurations is detected in the \ruv$=80$~\kl\ map, whereas this fraction 
in \irasA\ is only 25\%. Therefore, it appears that a  higher fraction of the \irasB\ 
envelope has already accreted into the disk and protostar than in \irasA.
This suggests that, as proposed by theory, the well formed \irasB\ 
magnetized disk is due to the fact that  \irasB\ is probably more evolved than \irasA. Further evidence supporting this argument stems from the lack of outflow activity in \irasB, when compared 
to \irasA\ \citep{rao09, Alves12, Girart13}.

Previous observations of circumstellar disks failed to detect the polarized dust
emission down to stringent upper limits, $\la 0.5$\%, \citep{Krejny09, Hughes09, 
Hughes13}. These upper limits are well below the $\simeq 2$-3\% polarization fractions expected
from simulations of a magnetized circumstellar disk (\ie\ having a 
predominantly toroidal magnetic field) with standard dust grain alignments
 \citep{Cho07}. These results have been interpreted either due to 
inefficient alignment of large grains, or insufficient magnetic field strength 
for alignment, or as a consequence of magnetic field tangling by turbulent motions
\citep{Hughes13}. 
The polarization fractions detected toward the \irasB\ disk are in the 0.5-3.0\%, 
with a median value of 1.4\%. These values are above the upper limits obtained 
previously in other YSOs, but slightly below the predicted values from the 
theory. For a better comparison with the expected polarization properties,
observations at higher angular resolution and sensitivity are definitively 
needed. It is worth noting that the sample used in previous observations 
are from YSOs significantly more evolved than \irasB, as they are classical T Tauri
or HAeBe stars \citep[HD 163296, TW Hydra, GM Aur, DG Tau and 
MWC 480:][]{Krejny09, Hughes09, Hughes13}. Thus, we speculate that the 
polarization efficiency decreases with time, either due to a loss of 
alignment efficiency or by other causes. 

In summary, we report the first detection of linearly polarized dust emission
in  a circumstellar disk around a young stellar object from SMA
observations made at submillimeter wavelengths.  The magnetic field configuration roughly resembles that from a
rotating disk with  the toroidal field lines being wrapped by the rotating gas
and dust, forming a spiral pattern. Clearly, the significant improvement in
the sensitivity and image quality that ALMA provides will allow us to better sample
the magnetic field geometry in the \irasB\ disk, and to map the magnetic 
field structure in the \irasA\ circumbinary disk.

\acknowledgments
The Submillimeter Array is a joint project between
the Smithsonian Astrophysical Observatory and the Academia Sinica
Institute of Astronomy and Astrophysics.
We thank all members of the
SMA staff that made these observations possible. JMG also thanks the SMA staff
at Hilo for their support. JMG is  supported by the Spanish MINECO
AYA2011-30228-C03-02 and Catalan AGAUR 2009SGR1172 grants. 
SPL acknowledges support from the National Science Council of Taiwan
with Grants NSC 98-2112- M-007- 007-MY3, NSC 101-2119-M-007-004, and
102-2119-M-007-004-MY3.

\end{document}